\documentclass[journal=apchd5,manuscript=article]{achemso}

\usepackage[version=3]{mhchem} \usepackage[T1]{fontenc}       
\usepackage{amsmath}
\usepackage{cancel}
\usepackage{amssymb}
\usepackage{graphicx}
\usepackage{braket}
\usepackage[dvipsnames]{xcolor}
\usepackage{epstopdf}
\usepackage{textcomp}
\usepackage{siunitx}
\usepackage{placeins}

\usepackage{xr}
\usepackage[final]{pdfpages}
\AppendGraphicsExtensions{.tif}

\newcommand\rr{\mathbf{r}}
\newcommand\Gp{\mathbf{G}_{+}(\rr)}
\newcommand\Gm{\mathbf{G}_{-}(\rr)}

\newcommand\Gppb{\mathbf{G}_{+}^{+}(\bar{\rr})}
\newcommand\Gmpb{\mathbf{G}_{-}^{+}(\bar{\rr})}
\newcommand\Gmmb{\mathbf{G}_{-}^{-}(\bar{\rr})}
\newcommand\Gpmb{\mathbf{G}_{+}^{-}(\bar{\rr})}

\newcommand\Gpp{\mathbf{G}_{+}^{+}(\rr)}
\newcommand\Gmp{\mathbf{G}_{-}^{+}(\rr)}
\newcommand\Gmm{\mathbf{G}_{-}^{-}(\rr)}
\newcommand\Gpm{\mathbf{G}_{+}^{-}(\rr)}

\newcommand\Gppbr{R_3\lb\mathbf{G}_{+}^{\mathrm{tot},+}\rb(\bar{\rr})}
\newcommand\Gppbrnt{R_3\lb\mathbf{G}_{+}^{+}\rb(\bar{\rr})}
\newcommand\Gmpbr{R_3\lb\mathbf{G}_{-}^{\mathrm{tot},+}\rb(\bar{\rr})}
\newcommand\Gmpbrnt{R_3\lb\mathbf{G}_{-}^{+}\rb(\bar{\rr})}
\newcommand\Gmmbr{R_3\lb\mathbf{G}_{-}^{\mathrm{tot},-}\rb(\bar{\rr})}
\newcommand\Gpmbr{R_3\lb\mathbf{G}_{+}^{\mathrm{tot},-}\rb(\bar{\rr})}

\newcommand\Gppt{\mathbf{G}_{+}^{\mathrm{tot},+}(\rr)}
\newcommand\Gmpt{\mathbf{G}_{-}^{\mathrm{tot},+}(\rr)}
\newcommand\Gmmt{\mathbf{G}_{-}^{\mathrm{tot},-}(\rr)}
\newcommand\Gmmtnt{\mathbf{G}_{-}^{-}(\rr)}
\newcommand\Gpmt{\mathbf{G}_{+}^{\mathrm{tot},-}(\rr)}
\newcommand\Gpmtnt{\mathbf{G}_{+}^{-}(\rr)}

\newcommand\lb{\left(}
\newcommand\rb{\right)}

\newcommand\Gip{{G}^{in}_{+}}
\newcommand\Gim{{G}^{in}_{-}}
\newcommand\Gop{{G}_+^{out}}
\newcommand\Gom{{G}_-^{out}}
\newcommand\Gopp{{G}_+^{out,+}}
\newcommand\Gomp{{G}_-^{out,+}}
\newcommand\Gopm{{G}_+^{out,-}}
\newcommand\Gomm{{G}_-^{out,-}}

\newcommand\EE{\mathbf{E}(\rr)}
\newcommand\HH{\mathbf{H}(\rr)}
\newcommand\EEmr{\mathbf{E}(-\rr)}
\newcommand\HHmr{\mathbf{H}(-\rr)}

\newcommand\equaldueto[1]{\stackrel{#1}{=}}

\newcommand{\onlinecite}[1]{\hspace{-1 ex} \nocite{#1}\citenum{#1}} 
\author{Florian Graf}
\affiliation[]{Institute of Theoretical Solid State Physics, Karlsruhe Institute of Technology, 76128 Karlsruhe, Germany}
\email{florian.graf.90@web.de}

\author{Joshua Feis}
\affiliation[]{Institute of Applied Physics, Karlsruhe Institute of Technology, 76128 Karlsruhe, Germany}

\author{Xavier Garcia-Santiago}
\affiliation[]{Institute of Nanotechnology, Karlsruhe Institute of Technology, 76021 Karlsruhe, Germany}
\alsoaffiliation[]{JCMWave GmbH, 14050 Berlin, Germany}

\author{Martin Wegener}
\affiliation[]{Institute of Applied Physics, Karlsruhe Institute of Technology, 76128 Karlsruhe, Germany}
\alsoaffiliation[]{Institute of Nanotechnology, Karlsruhe Institute of Technology, 76021 Karlsruhe, Germany}

\author{Carsten Rockstuhl}
\affiliation[]{Institute of Theoretical Solid State Physics, Karlsruhe Institute of Technology, 76128 Karlsruhe, Germany}
\alsoaffiliation[]{Institute of Nanotechnology, Karlsruhe Institute of Technology, 76021 Karlsruhe, Germany}
\author{Ivan Fernandez-Corbaton}
\affiliation[]{Institute of Nanotechnology, Karlsruhe Institute of Technology, 76021 Karlsruhe, Germany}
\email{ivan.fernandez-corbaton@kit.edu}

\title[An \textsf{achemso} demo]
  {Achiral, Helicity Preserving, and Resonant Structures for Enhanced Sensing of Chiral Molecules}

\abbreviations{}

\begin{document}

\begin{abstract}
	We derive a set of design requirements that lead to structures suitable for molecular circular dichroism (CD) enhancement. Achirality of the structure and two suitably selected sequentially incident beams of opposite helicity ensures that the CD signal only depends on the chiral absorption properties of the molecules, and not on the achiral ones. Under this condition, a helicity preserving structure, which prevents the coupling of the two polarization handednesses, maximizes the enhancement of the CD signal for a given ability of the structure to enhance the field. When the achirality and helicity preservation requirements are met, the enhancement of the CD signal is directly related to the enhancement of the field. Next, we design an exemplary structure following the requirements. The considered system is a planar array of silicon cylinders under normally incident plane-wave illumination. Full-wave numerical calculations show that the enhancement of the transmission CD signal is between 6.5 and 3.75 for interaction lengths between 1.25 and 3 times the height of the cylinders.
\end{abstract}

\section*{}

Chiral objects are omnipresent in and all around us, ranging from the double-helical structure of our DNA, extending over our hands - that originally lent the property its name - up to the spiral galaxies in the sky. Chiral objects, which are not super-imposable onto their mirror images by any translation or rotation, play a fundamental role in modern science as well as life itself. The reason for the homochirality of living nature is still one of its greatest secrets \cite{Quack2002}. Apart from the fundamental questions arising due to the violation of mirror symmetry in the laws of our universe, we often deal with more pragmatic problems. Enantiomers, being pairs of mirror-image chiral molecules, often react differently in biological organisms due to the chiral specifications of the cells' receptors. As a result, drugs consisting of chiral molecules can have profoundly different therapeutic and/or toxicological properties \cite{Nguyen2006}. Sharing the same atomic composition, pairs of enantiomers are indistinguishable when measuring their scalar physical properties. Their chiral nature is mostly revealed in the interaction with other chiral objects. In optics, the most common chiral object that we deal with is circularly polarized light (CPL). Upon interaction with light, a chiral molecule generally exhibits a preferential absorption for either left or right circular polarization. This difference is measured by means of circular dichroism (CD) spectroscopy. In a traditional CD setup, the molecular solution is sequentially illuminated by propagating beams of different circular polarization handedness, and the total transmitted power of light is recorded in each case. The CD signal is the difference between the two power measurements. The differential absorption is due to the chiral absorption properties of the molecule and has the opposite sign for the two enantiomers. The chiral differential absorption is typically orders of magnitude smaller than the achiral molecular absorption, which is the same for both enantiomers. For samples of low molecular concentration, limits concerning illumination power lead to long measurement times, up to several hours \cite{Polavarapu2004}, that are needed in order to reliably detect the CD signal over the noise. Moreover, some envisioned applications, e.g. in the context of lab-on-a-chip, require the testing of minute quantities of analytes in small volumes that are not easily accessed by a focused propagating beam. High field intensities in confined volumes are hence required. This need can be met by photonic structures that resonantly enhance the near-field. By following the paradigm of the typical CD experiment we can intuitively understand that a high field enhancement is not the only desired feature. It is intuitively clear that the near-fields should be of pure handedness. A near-field of mixed handedness is similar to a traditional CD setup with the two beams having mixed polarization handedness: For a fixed illumination power, the CD signal would decrease. Additionally, since we want to probe a chiral property of the molecules, it seems appropriate to avoid any chirality in the structures, thereby maintaining the parallelism with CD setups without any structure. On the one hand, if a chiral structure has absorption, it will introduce its own CD signal. The distinction of this usually strong CD signal from the much weaker molecular CD signal requires an accurate calibration in order to not compromise the detection of the actual molecular signal. Additionally, even if the structure is lossless, its chiral character may result in different near-field enhancements upon illumination with beams of different handedness but equal power. This difference will bias the molecular CD measurement as well. Many metallic and dielectric structures, resonant and non-resonant, have been studied for the purpose of enhancing the CD signal \cite{Auguie2011,Schaeferling2012,Hentschel2012,Valev2013,GarciaEtxarri2013,Nesterov2016,GarciaEtxarri2017,Zhao2017,Vazquez-Guardado2018,Hanifeh2018b,Mohammadi2018,Garcia-Guirado2018}. In most of these structures, the requirement of pure handedness in the near-fields is not fullfilled. For example, metallic structures can produce large near-field enhancements due to the excitation of plasmonic resonances, but very often lack the ability to preserve the polarization handedness (helicity) of the illumination. This leads to hot spots of both helicities in the resulting near-fields. Notably, recent work shows that metal-based structures are still able to meet the requirement of helicity preservation by carefully designing them in order to overlap ``magnetic'' and ``electric''-like modes \cite{Vazquez-Guardado2018}. The effectiveness of this electric/magnetic balance can be traced back to the electromagnetic duality symmetry of a scatterer, which, by definition, preserves the helicity of the illuminating field and hence allows for near-fields of pure handedness (see e.g. Fig. 1 in Ref. (\onlinecite{FerCor2012p})). Other recent proposals exploit the off-resonance helicity preservation properties of homogeneous dielectric spheres \cite{Zambrana2016b,Hanifeh2018b}. Besides the inability to preserve the helicity in the near-field, some of the proposed structures are chiral, which leads to the aforementioned problems. For systems based on non-mirror symmetric unit cells, chirality has been avoided, for example, by interaction with the two mirror versions of a chiral array \cite{Zhao2017}, or by combining the two asymmetric unit cells onto the same structure \cite{Schaeferling2012,Garcia-Guirado2018}. Other solutions use mirror-symmetric structures \cite{GarciaEtxarri2013,Nesterov2016,Zambrana2016b,GarciaEtxarri2017,Mohammadi2018,Vazquez-Guardado2018,Hanifeh2018b}. In Ref. (\onlinecite{Tang2011}), a lossy mirror was used to create a standing-wave field in the molecular solution. Such system meets the achirality condition but does not meet the helicity-preserving condition since a mirror reflection changes the helicity of the reflected beam with respect to that of the incident beam. The idea of a cavity-like system is appealing as a proven way to enhance the light-matter interaction. Nevertheless, the configuration in Ref. (\onlinecite{Tang2011}) could be improved by the use of a helicity preserving mirror. Unfortunately, the current designs for such devices (see, e.g. Ref. (\onlinecite{Plum2015})) are chiral themselves.

In this work, we analytically show the importance of using \textit{achiral and helicity-preserving structures with a strong electromagnetic response} to significantly increase the CD signal obtained from an absorbing chiral environment. If the structure has at least one space-inversion symmetry, which makes it achiral, and the two sequentially illuminating beams transform into each other through that symmetry operation, it is guaranteed that the differential signal only depends on the chiral absorption properties of the molecules, and not on the achiral ones. Under this condition, helicity preservation maximizes the CD enhancement for a given field enhancing ability of the structure. When the two previous requirements are met, the CD enhancement is directly related to the strength of the electromagnetic response of the structure. Next, we design an exemplary structure following these requirements. The considered system is a planar array of silicon cylinders connected by thin rods under normally incident plane-wave illumination. The reflection symmetry of the arrays across some planes containing the optical axis, under which the two incident circularly polarized plane-waves transform into each other, ensures that the CD signal only depends on the chiral absorption properties of the molecules. The aspect ratio of the cylinders is tuned such that, for normally incident plane-waves, helicity is preserved at a given excitation wavelength in the vicinity of the dipolar resonances. The resulting near-field is of high intensity \textit{and, crucially}, maintains the helicity of the illumination to a high degree. Full-wave numerical calculations show that the enhancement of the transmission CD signal is between 6.5 and 3.75 for interaction lengths between 1.25 and 3 times the height of the cylinders. Compared to the case where the array is absent, the enhancement would provide the same several-fold reduction of the measurement integration time for a given target signal to noise ratio, or the same several-fold increase in the signal-to-noise ratio for a fixed integration time.

In the rest of the article, we start by formalizing the design requirements. Then, we show that silicon cylinders can be tailored to meet the requirements, first using a single cylinder and later considering planar arrays of cylinders, both free-standing and connected by thin rods. Finally, we evaluate the CD enhancement of the connected array. After introducing the problem using the most common formalism, our analysis is based on the Riemann-Silberstein representation of Maxwell's fields \cite{Birula1996,Birula2013}, and the advantages over the electric and magnetic field representation in this context are highlighted. We assume monochromatic fields with a $\exp(-\text{i}\omega t)$ dependence throughout the article.

\paragraph{Design Requirements for CD Enhancing Structures:} The interaction between the electromagnetic field and a sufficiently small object, such as a molecule or a nanostructure, can be treated in the linear dipolar approximation, which features a 6$\times$6 polarizability tensor that relates the external electric and magnetic fields $\{\textbf{E}(\rr),\textbf{H}(\rr)\}\in \mathbb{C}^3$ at the position of the object, to the induced electric and magnetic dipole moments $\textbf{p}(\rr)$ and $\textbf{m}(\rr)$:
\begin{align}
	\label{eq:dipoles}
\left(\begin{array}{c} \textbf{p}(\rr) \\ \textbf{m}(\rr) \end{array}\right) =\left(\begin{array}{rr} \underline{\underline{\alpha}}_\text{ee} & \underline{\underline{\alpha}}_\text{em} \\
      \underline{\underline{\alpha}}_\text{me} & \underline{\underline{\alpha}}_\text{mm} \end{array}\right)\left(\begin{array}{c} \textbf{E}(\rr) \\ \textbf{H}(\rr) \end{array}\right),
\end{align}
where the $\underline{\underline{\alpha}}_{\text{ab}}$ are 3$\times$3 tensors. We now consider the absorption of a randomly oriented molecule at position $\rr$, immersed in a lossless medium characterized by electric permittivity $\epsilon$ and magnetic permeability $\mu$. The rotational average of the single-molecule absorption for different orientations is often approximated by the absorption of a rotationally averaged polarizability tensor, where all the $\underline{\underline{\alpha}}_{\text{ab}}$ turn into scalars. For scalars, we have that $\alpha_{\text{em}}=-\mu\alpha_{\text{me}}$ because of reciprocity\cite[Eq. 7 and Tab. I]{Sersic2011}. If we furthermore assume that the field scattered by the molecule is much smaller than the external incident field, we can quantify the absorption of light by a randomly oriented chiral molecule located at a position $\rr$ by 
\begin{align}
	\label{eq:A}
	A(\rr) = \underbrace{\frac{\omega}{2}\left(\alpha_{\text{ee}}''|\textbf{E}(\rr)|^2+\mu\alpha_{\text{mm}}''|\textbf{H}(\rr)|^2\right)}_{\text{achiral absorption } f(\alpha_\text{ee}'',\alpha_\text{mm}'')}\underbrace{-2c^2\alpha_{\text{em}}'C(\rr)}_{\text{chiral absorption }g(\alpha_\text{\text{em}}')},
\end{align}
where $c=1/\sqrt{\epsilon\mu}$, $\alpha_{\text{ab}}^{''}$ denotes imaginary part of $\alpha_{\text{ab}}$ and $\alpha_{\text{ab}}^{'}$ its real part. $C(\mathbf{r})=-\frac{\epsilon\omega}{2}\text{Im}\{\mathbf{E}(\rr)^\dagger\mathbf{B}(\rr)\}$ is the optical chirality density, a measure for the chirality density of the electromagnetic field \cite{Tang2010}. We observe that the absorption can be decomposed into a part that depends on the achiral absorption properties of the molecule, $\alpha_{\text{ee}}''$ and $\alpha_{\text{mm}}''$, and a part that depends on the chiral absorption of the molecule, $\alpha_{\text{em}}'$. The reason for the small differential absorption of chiral molecules lies in the intrinsically small values of $\alpha_{\text{em}}'$ in Eq. (\ref{eq:A}), which typically renders the chiral absorption much weaker than the achiral absorption \cite{Barron2004}. 

We now set out to formalize requirements for the design of structures aiming at enhancing the CD signal. We start by changing the representation of the electromagnetic field to the Riemann-Silberstein vectors \cite{Birula1996,Birula2013}:
\begin{align}
	\label{eq:rs}
	\textbf{G}_\pm(\rr)=\textbf{E}(\rr)\pm \text{i}Z \textbf{H}(\rr),
\end{align}
which are the eigenstates of the helicity operator with eigenvalues $\pm$1. In a field that can be decomposed into propagating and/or evanescent plane-waves that are exclusively left (right) circularly polarized, $\textbf{G}_-(\rr)$ ($\textbf{G}_+(\rr)$) vanishes at all space-time points \cite[Chap. 2]{FerCorTHESIS}. In fields that contain both polarization handednesses, both $\textbf{G}_+(\rr)$ and $\textbf{G}_-(\rr)$ are non-zero. 

One of the advantages of using $\textbf{G}_\pm(\rr)$ in the present context is the connection between the helicity eigenstates and the optical chirality density \cite[Eq. (5)]{FerCor2016}
\begin{equation}
\label{eq:C}
	C(\rr)=-\frac{\epsilon\omega}{2}\text{Im}\{\mathbf{E}(\rr)^\dagger\mathbf{B}(\rr)\}=\frac{\epsilon\omega}{8c}\left[|\textbf{G}_+(\rr)|^2-|\textbf{G}_-(\rr)|^2\right],
\end{equation}
which reveals an arguably more intuitive relationship than using $\text{Im}\{\mathbf{E}(\rr)^{\dagger}\mathbf{B}(\rr)\}$: The local optical chirality density is proportional to the difference between the local norm square of the two helicity eigenstates. Using Eqs. (\ref{eq:rs})-(\ref{eq:C}) and straightforward manipulations, we can write Eq. (\ref{eq:A}) as:
\begin{equation}
	\begin{split}
		A(\rr) &= \frac{\omega}{8}\left[\lb |\Gp|^2+|\Gm|^2\rb\lb\alpha_{\text{ee}}''+\epsilon\alpha_{\text{mm}}''\rb+2\mathrm{Re}\{{\Gp}^\dagger\Gm\}\lb \alpha_{\text{ee}}''- \epsilon\alpha_{\text{mm}}''\rb\right]\\
		&-\frac{\omega}{4Z}\lb |\Gp|^2-|\Gm|^2\rb\alpha_{\text{em}}'.
	\end{split}
\end{equation}
In a CD measurement, the system containing the molecular solution is subsequently illuminated by two beams of opposite helicity. The difference between the total outgoing power in the two cases is the CD signal, which will be the integral of the differential absorption in the volume occupied by the molecules. If we use $A^{\pm}(\rr)$ to denote the point-wise absorption upon illumination with the two beams, and $\mathbf{G}_\lambda^{\bar{\lambda}}(\rr)$ to denote the component of helicity $\lambda$ at point $\rr$ due to an external beam of helicity $\bar{\lambda}$, we get:
\begin{equation}
	\label{eq:datext}
	\begin{split}
		\Delta A (\rr)&=A^+(\rr)-A^-(\rr)=\\
		&\frac{\omega}{8}\lb |\Gpp|^2+|\Gmp|^2-|\Gmm|^2-|\Gpm|^2\rb\lb\alpha_{\text{ee}}''+\epsilon\alpha_{\text{mm}}''\rb\\
		&+\frac{\omega}{4}\lb \mathrm{Re}\{{\Gpp}^\dagger\Gmp-{\Gpm}^\dagger\Gmm\}\rb\lb\alpha_{\text{ee}}''-\epsilon\alpha_{\text{mm}}''\rb\\
		&-\frac{\omega}{4Z}\lb |\Gpp|^2-|\Gmp|^2-|\Gpm|^2+|\Gmm|^2\rb\alpha_{\text{em}}'. 
	\end{split}
\end{equation}
We note that the first two lines depend on the achiral absorption properties of the molecules, and only the third line on their chiral absorption properties. The CD measurement system must ensure that the contribution of the first two lines integrated over the volume occupied by the molecules cancels out. Without this cancellation, the achiral absorption would blur or even dominate the desired signal. Moreover, achiral molecules would produce a CD signal. Additionally, we want to design CD measurement systems that enhance the integrated value of the third line. The Supporting Information contains the detailed analytical derivations that address these requirements. In what follows, we summarize the results. 

The first result is that the cancellation of the terms that depend on the achiral absorption properties of the molecules $(\alpha_{\text{ee}}'',\alpha_{\text{mm}}'')$ is {\em guaranteed} by using two incident fields of pure and opposite helicity that transform into each other by means of a space-inversion operation that is a symmetry of the structure. That means that the structure has at least one space-inversion symmetry, which makes it achiral, and that the two illumination beams are selected accordingly. The cancellation of the integrated contributions from the first two lines of Eq. (\ref{eq:datext}) happens when the differential absorption at each point $\rr$ is added to the one at another point $\bar{\rr}$ which is the image of $\rr$ after the mentioned space-inversion operation.

In light of the above result, we take achirality as a requirement for the structure, and assume that the illuminations have the necessary relationship. Under this conditions, the chirality density is a good proxy for the CD signal, and maximizing it also maximizes the CD signal. Indeed, the CD signal can be then written [Eqs. (\ref{eq:dadabar}) and (\ref{eq:cisgood}) in the Supporting Information]:
\begin{equation}
	\label{eq:dadabartext}
	\begin{split}
		\text{CD}&=-\frac{\alpha_{\text{em}}'\omega}{4Z}\int_V d\rr\left[ \lb|\Gpp|^2-|\Gmp|^2\rb-\lb|\Gpm|^2-|\Gmm|^2\rb\right]\\
		&\equaldueto{\text{Eq. (\ref{eq:C})}}-2\alpha_{\text{em}}'c^2\int_V d\rr \left[ C^+(\rr)-C^-(\rr)\right]\\
		&=-4\alpha_{\text{em}}'c^2\int_V d\rr \ C^+(\rr)=4\alpha_{\text{em}}'c^2\int_V d\rr \ C^-(\rr).
	\end{split}
\end{equation}
where $C^\lambda(\rr)$ is the optical chirality density at point $\rr$ when the system is illuminated with the beam of helicity $\lambda$. Let us now consider the maximization of the CD signal given a total integrated field intensity, which is proportional to 
\begin{equation}
	\label{eq:power}
	\int_V d\rr \left[\lb|\Gpp|^2+|\Gmp|^2\rb+\lb|\Gmm|^2+|\Gpm|^2\rb\right],
\end{equation}
and depends on the power of the incident beams and on the field enhancing ability of the structure. Assuming a fixed incident beam power, and considering Eqs.\ (\ref{eq:dadabartext}) and (\ref{eq:power}), it becomes clear that, {\em for a given total field enhancement, a helicity preserving structure, which will force $\Gmp=\Gpm=0$ for all $\rr$, will maximize the CD enhancement.} We note that a non-helicity preserving structure can outperform a helicity preserving structure through a larger field enhancing ability.

We hence take helicity preservation as a requirement for the design of the structure. 

The benefits of helicity preservation can be intuitively grasped from the ratio of optical chirality density $C(\rr)$ and energy density of the field $U(\rr)$:
\begin{align}
	\frac{C(\rr)}{U(\rr)}=\frac{\frac{\epsilon\omega}{8c}\left[|\textbf{G}_+(\rr)|^2-|\textbf{G}_-(\rr)|^2\right]}{\frac{\epsilon}{8c}\left[|\textbf{G}_+(\rr)|^2+|\textbf{G}_-(\rr)|^2\right]}=\frac{\omega}{c}\frac{|\textbf{G}_+(\rr)|^2-|\textbf{G}_-(\rr)|^2}{|\textbf{G}_+(\rr)|^2+|\textbf{G}_-(\rr)|^2}\,
\label{C/U}
\end{align}
from which one can deduce that, for a fixed energy density, the optical chirality values are extremal when either $\textbf{G}_+(\rr)$ or $\textbf{G}_-(\rr)$ vanishes, i.e. when the field is of pure helicity. 

When the structure preserves helicity upon illumination with the two incident beams, the CD signal can be written by setting $\Gmp=\Gpm=0$ in the first line of Eq. (\ref{eq:dadabartext}) as:
\begin{equation}
	\label{eq:CDs}
	\text{CD}=-\frac{\alpha_{\text{em}}'\omega}{4Z}\int_V d\rr\left[ |\Gpp|^2+|\Gmm|^2\right]=-\frac{\alpha_{\text{em}}'\omega}{2Z}\int_V d\rr |\Gpp|^2=-\frac{\alpha_{\text{em}}'\omega}{2Z}\int_V d\rr |\Gmm|^2,
\end{equation}
where the second and third equalities follow from the relationship between the $\Gpp$ at point $\rr$ and the $\mathbf{G}_-^{-}(\bar{\rr})$ field at the image point $\bar{\rr}$ [see the Supporting Information and its Eq. (\ref{eq:equalities})]. Equation (\ref{eq:CDs}) makes clear that, besides achirality and helicity preservation, a strong response of the structure is needed in order to get a large field enhancement and increase the CD signal. It should be noted that the volume occupied by the structure is lost to the molecules. Therefore, compared to the case without structure, the field enhancement must first compensate for this reduction in analyte volume before it can produce a net gain.

Let us now further discuss the requirements. Designing structures with at least one space-inversion symmetry, and selecting the illuminating fields of pure helicity so that such symmetry transformation maps them into each other is a conceptually simple endeavor. For example, the structures that we will consider later, isolated cylinders and cylinder arrays, have several planes of reflection symmetry that include the rotational symmetry axis of the cylinders. Then, there is a large class of illuminating beams propagating along such an axis from where appropriate pairs of beams with opposite helicity can be selected: Plane-waves, Gaussian beams, and also vortex beams \cite{Zambrana2016}. With regards to large field enhancements, such requirement is often successfully addressed by designing resonant structures.

The requirement of helicity preservation is arguably more challenging. In general, structures do not conserve the polarization handedness of the illumination during the scattering process. That is, the interaction couples the $\mathbf{G}_{\pm}(\rr)$ components. The near-field of a general structure illuminated with a field of positive(negative) helicity also contains negative(positive) helicity components, sometimes in a roughly equal mixture. The conversion of helicity in the presence of matter is controlled by a conservation law \cite{Nienhuis2016}, and lends itself to quantification \cite{Nieto2015,Poulikakos2016,Gutsche2016,FerCor2016c,Vazquez2018}. The symmetry connected to this conservation law, which ensures helicity preservation, is the electromagnetic duality symmetry. It can be understood as the interchangeability of electric and magnetic properties. This symmetry is typically broken by the presence of matter. Duality symmetry, and therefore also helicity preservation \textit{for any illumination}, is restored in a geometry independent way by the condition that the ratio of permittivity to permeability is the same for all materials $\epsilon_\text{r}(\omega)/\mu_\text{r}(\omega)=\text{const}$\cite{FerCor2012p}. The fact that natural materials do not show a significant magnetic response in many frequency bands (e.g., in the optical regime $\mu_\text{r}\approx 1$) prevents us very often from achieving helicity preserving structures in this way. However, when objects can be treated as dipolar scatterers, they can also be made helicity preserving in that approximation by choosing appropriate combinations of non-magnetic materials and geometry \cite{Zambrana2013b,FerCor2015,FerCor2016,Rahimzadegan2017}. To obtain a dual symmetric dipolar scatterer, the tensors in Eq. (\ref{eq:dipoles}) must meet $\underline{\underline{\alpha}}^\text{ee}=\epsilon\underline{\underline{\alpha}}^\text{mm}$ and $\underline{\underline{\alpha}}^\text{em}=-\mu\underline{\underline{\alpha}}^\text{me}$ (see e.g. Refs. (\onlinecite{Karilainen2012,FerCor2013,Nieto2015})). Such a particle will preserve the helicity for {\em any illumination}, that is, it will never couple the incident $\mathbf{G}_\pm(\rr)$. Each of the $\mathbf{G}_\pm(\rr)$ will independently excite electric and magnetic dipole moments in the object related by $\textbf{p}=\pm \frac{\text{i}}{c}\textbf{m}$. Such combinations are pure helical dipoles that radiate dipolar waves of well defined $\pm$1 helicity \cite{FerCor2013}. The same principle applies to objects that are large enough to have a significant response of higher multipolar order: The electric and magnetic quadrupoles need to be equal (in appropriate units). The condition of dipolar duality, or duality in general, can be relaxed when the illumination conditions are known \textit{a priori} to a good approximation. It is then possible to design particles with a high degree of helicity preservation {\em upon prescribed illuminations}. This is the setting that we exploit in our article.

We now follow the requirements to design a particular structure for CD enhancement in the infrared. We will first optimize the aspect ratio of a silicon cylinder so that the electric and magnetic responses align themselves in a near-resonant way that, besides enhancing the field, preserves the incident helicity to a high degree \cite{Evlyukhin2011,Staude2013,Chong2016}. The cylinders are then arranged periodically in a square array, and connected with small rods to obtain a connected structure without the need for an additional substrate. After a further incremental optimization we demonstrate good helicity preserving and optical chirality enhancement properties of the array. Finally, we calculate the enhancement of circular dichroism in transmission obtained from a solution of chiral molecules (modeled as a Pasteur medium) surrounding the array. We highlight that the mirror symmetries of the connected cylinder array, which transform the incident left-handed plane-wave into the incident right handed plane-wave, ensure that the CD signal only depends on the chiral absorption properties of the molecules ($\alpha_{\text{em}}'$), and not on their achiral absorption properties ($\alpha_{\text{ee}}''$, and $\alpha_{\text{mm}}''$).
\section*{Results and discussion}
\FloatBarrier
\begin{figure}[h!]
\includegraphics[width=\textwidth]{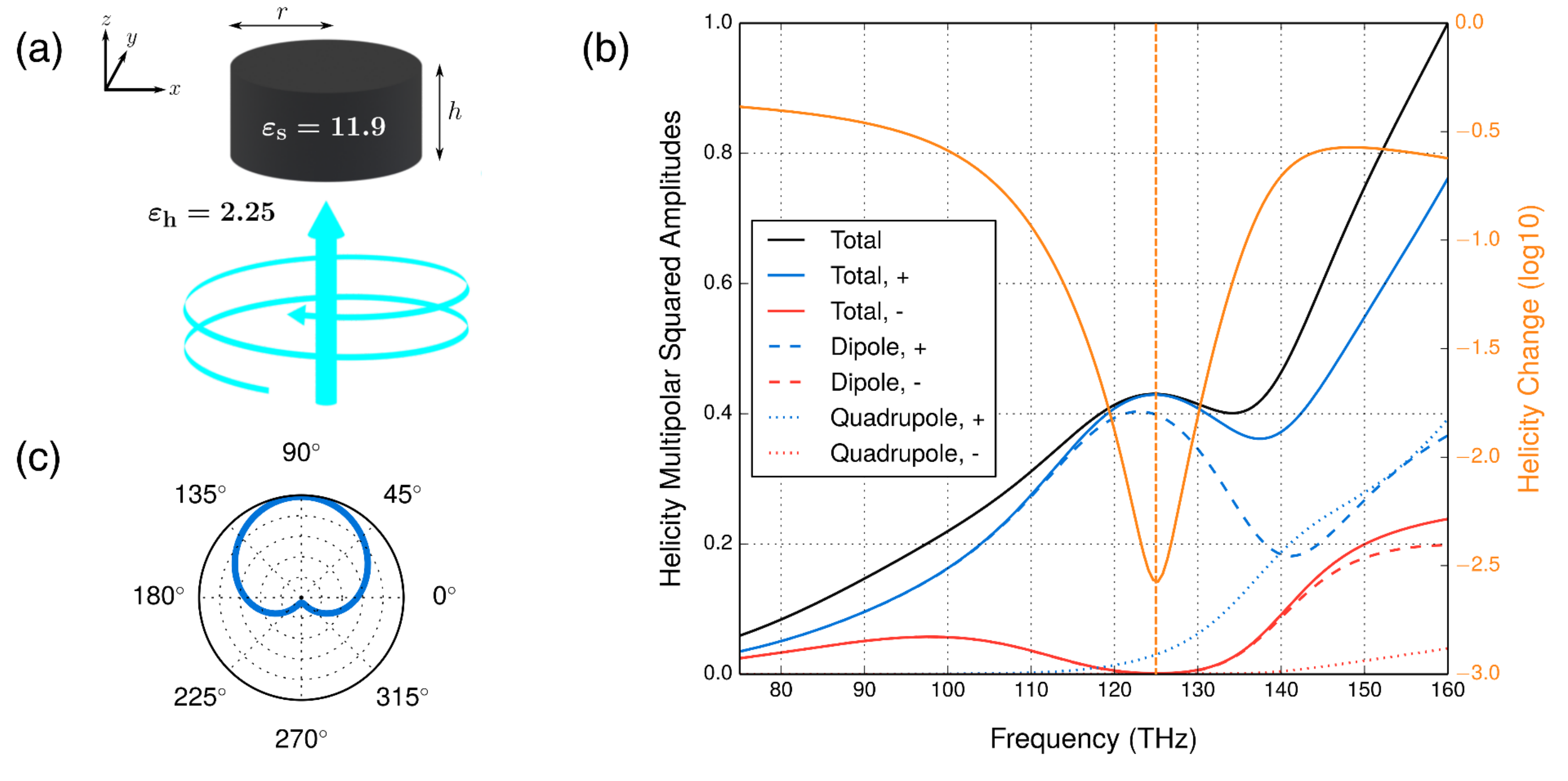}
	\caption{(a) Sketch of the considered scattering scenario. The silicon cylinder embedded in an $\varepsilon_\mathrm{h}=2.25$ medium is illuminated by a left-hand polarized plane-wave propagating in the $z$-direction. The radius of the cylinder is $r = \SI{448}{nm}$, the height is $h = \SI{400}{nm}$. (b) Right axis: Helicity change [$\cancel{\Lambda}$ in Eq. (\ref{eq:cl})] in logarithmic scale. Left axis: Different sums of the squared multipolar helicity amplitudes of the scattered field resulting from the interaction between the plane-wave and the cylinder. (c) Corresponding far-field radiation diagram in a polar plot.}
\label{Fig1}
\end{figure}
\paragraph{Single Cylinder:} A cylinder surrounded by an homogeneous medium of relative permittivity $\varepsilon_\text{h}=2.25$ is illuminated with a left-hand polarized plane-wave propagating along the axis of the cylinder, as depicted in Fig.\,\ref{Fig1} (a). The material of choice is silicon, whose high permittivity allows particles to host electric and magnetic dipolar resonances of comparable strength \cite{Evlyukhin2010,GarciaEtxarri2011,Kuznetsov2016}. The cylinder geometry is chosen in part because, while giving up isotropy, cylinders provide the ability to tune the relative spectral positions of the electric and magnetic resonances by changing the aspect ratio $r/h$ \cite{Evlyukhin2011,Staude2013,Chong2016}. This makes it possible to bring them into spectral overlap, resulting in a strong overall response, which additionally fulfills the dipole polarizability conditions for helicity preservation at normal incidence. We choose the near infrared region because it hosts many vibrational resonances of chiral molecules and proteins. The optimization is done by fixing the cylinder height to 400 nm, and varying the radius and the frequency. Inside the $2.25$ medium, the cylinder has a large enough electromagnetic size so that its quadrupolar response cannot be neglected. We decompose the scattered fields into multipoles of well defined helicity (see Methods). At each frequency, the helicity multipolar coefficients $g_{jm}^{\pm}(\omega)$ are readily obtained \cite[Eq. 11.4-25]{Tung1985}\cite[p. 18]{Berestetskii1982} from the electric $a_{jm}(\omega)$ and magnetic $b_{jm}(\omega)$ multipolar coefficients as $g_{jm}^{\pm}=\left[a_{jm}(\omega)\pm b_{jm}(\omega)\right]/\sqrt{2}$, where the $j=1,2,\ldots,$ index refers to the total angular momentum of the multipole, and the $m=-j\ldots j$ index to the angular momentum along the $z$ axis. It is sufficient to consider only $j=1$ (dipoles) and $j=2$ (quadrupoles) because the higher order terms are negligible. Since the incident plane-wave has helicity +1, the ratio 
\begin{equation}
	\label{eq:cl}
	\cancel{\Lambda}=\frac{\sum_{j,m} |g^-_{jm}|^2}{\sum_{j,m} |g^+_{jm}|^2+|g^-_{jm}|^2}
\end{equation}
indicates the degree of helicity change in the interaction. A similar measure was introduced in Ref. (\onlinecite{Zambrana2013b}). Figure \ref{Fig1}(b) shows $\cancel{\Lambda}$ in logarithmic scale for a cylinder of radius $r=448$ nm and height $h=400$ nm. We observe a pronounced dip in the helicity change at $f=\SI{125}{THz}$. Figure \ref{Fig1}(b) also shows, in linear scale, the sum of the squared amplitudes of: All coefficients, all helicity $\pm$1 coefficients, the helicity $\pm$1 dipoles, and the helicity $\pm 1$ quadrupoles. The minimum of $\cancel{\Lambda}$ is at $\SI{125}{THz}$. It coincides with a local maximum of the total sum, which is practically equal to the total helicity +1 sum at that frequency. The minimum of $\cancel{\Lambda}$ is also in the vicinity of the local maximum (resonance) of the helicity +1 dipole. The deviation from the dipolar resonance is due to a non-negligible contribution of the helicity +1 quadrupole. Figure \ref{Fig1} (c) shows the far field scattering pattern, and reveals that the energy is scattered mainly in the forward half space, and that there is practically zero energy in the specular reflection direction. This behavior is characteristic of helicity preserving objects with discrete rotational symmetries: Helicity preservation and a discrete rotational symmetry $2\pi/n$ for $n\ge 3$ are sufficient conditions for zero back-scattering \cite{FerCor2013c}. Square arrays of optimized silicon cylinders have been reported showing almost zero reflection \cite{Staude2013,Chong2016}. They are an example of $n=4$. Cylindrical symmetry is the case $n\rightarrow \infty$, which was initially studied by Kerker \cite{Kerker1983} using isolated spheres, and later generalized to any object of cylindrical symmetry \cite{Zambrana2013}. The potential for experimental realizations of zero backscattering with isolated spheres of high permittivity was recognized in Ref. (\onlinecite{Nieto2011}) and realized in Ref. (\onlinecite{Geffrin2012}), while other experimental demonstrations have used isolated cylinders \cite{Person2013}. 

We now analyze the helicity of the near-fields around the cylinder. We also analyze the corresponding enhancement of optical chirality which, as can be seen in Eq. (\ref{eq:dadabartext}), is a good proxy for the CD enhancement {\em if the structure is achiral and the two illuminations are related by a space-inversion symmetry of the structure}.
\FloatBarrier
\begin{figure}[h!]
\includegraphics[width=5.0in]{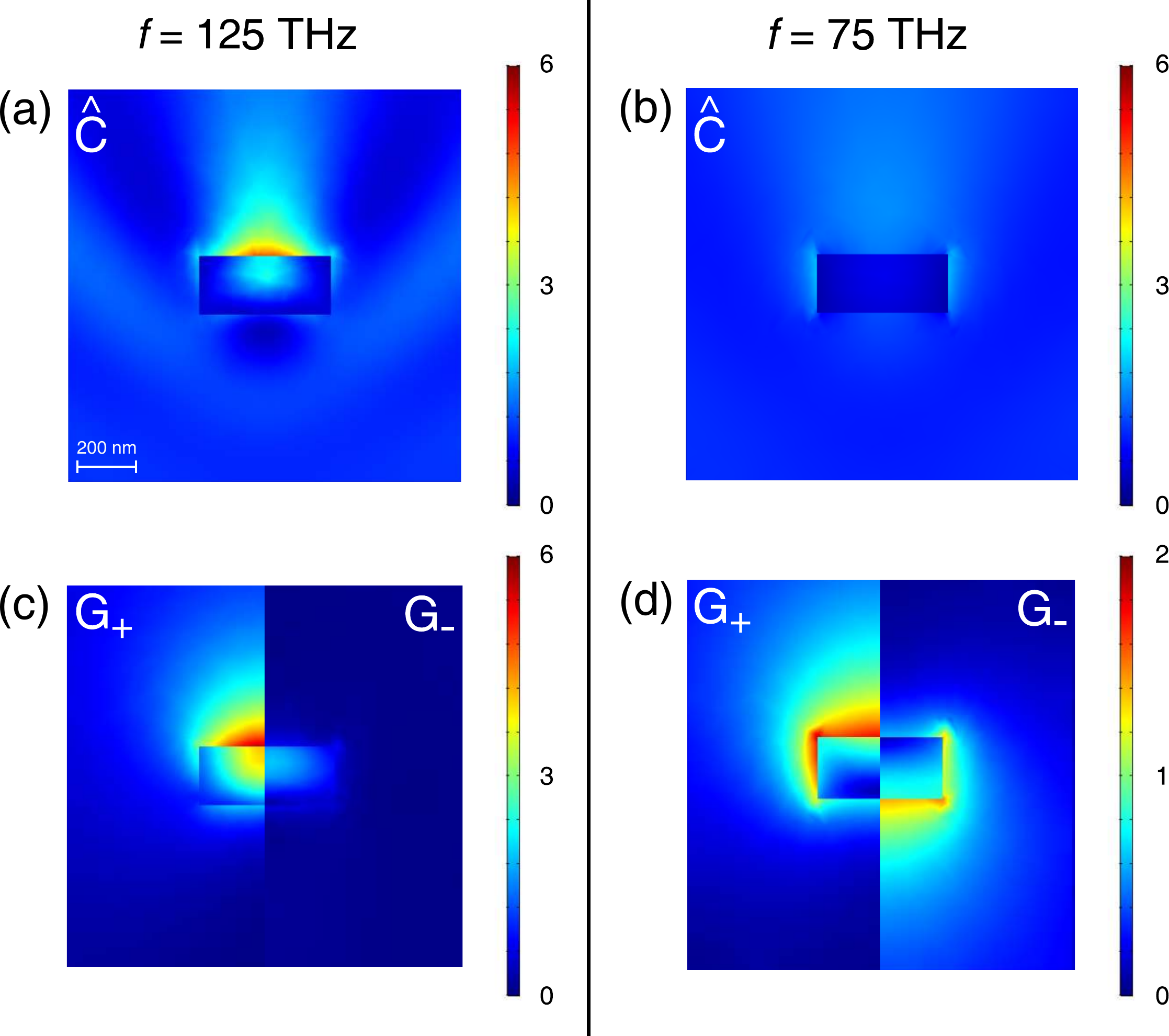}
	\caption{Panels (a) and (c) correspond to $f=\SI{125}{THz}$, and panels (b) and (d) to $f=\SI{75}{THz}$. Panels (a) and (b): Cross-section of the optical chirality density of the {\em total} electromagnetic field, normalized to the optical chirality density of the illuminating left-handed polarized plane-wave $|C_{\text{CPL}}|=\epsilon_{\text{h}}\omega/(2c)|E_0|^2$: $\hat{C}=C^+/|C_{\text{CPL}}|$. The parameters of the system are the same as in Fig. \ref{Fig1}. Panels (c) and (d): Cross sectional comparison of $|\textbf{G}_\pm(\rr)|=|\textbf{E}(\rr)\pm \text{i}Z\textbf{H}(\rr)|$ of the {\em scattered} field. }
\label{Fig2}
\end{figure}

\FloatBarrier
\paragraph{Helicity Preservation and Optical Chirality Enhancement:} The cylinder is illuminated with left circularly polarized light at the near-resonant and helicity preserving frequency of $\SI{125}{THz}$, and at $\SI{75}{THz}$, which corresponds to a non-resonant non-helicity preserving case. A low degree of helicity conversion in the scattered fields can be visually appreciated at $\SI{125}{THz}$ in Fig. \ref{Fig2}(c), and a roughly equal mixture of the scattered $\mathbf{G}_\pm(\rr)$ at $\SI{75}{THz}$ in Fig. \ref{Fig2}(d). This is fully consistent with Fig. \ref{Fig1}(b). The enhancement of the optical chirality density {\em for the total field} (incident plus scattered) is shown in Figs.\,\ref{Fig2}(a) and \ref{Fig2}(b). As expected, the optical chirality enhancement at $\SI{125}{THz}$ is considerably stronger than at $\SI{75}{THz}$. From these results, it can be concluded that the cylinder meets the three design requirements, and can act as an approximately helicity preserving source of chiral near-fields for enhanced CD measurements.
\begin{figure}[h!]
\includegraphics[width=\textwidth]{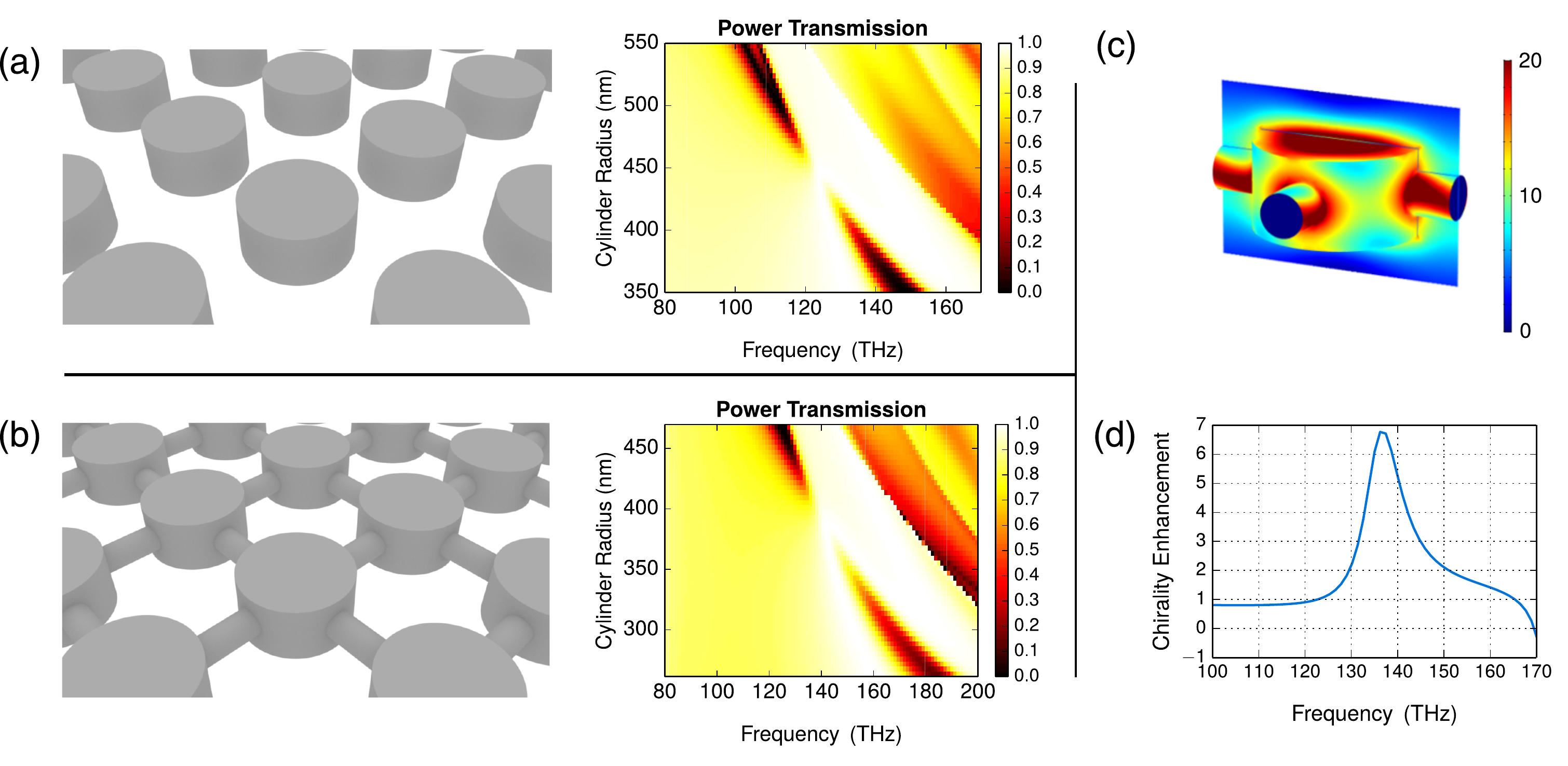}
	\caption{(a),(b) Depiction and power transmission spectra of 2D-periodic square lattices made of silicon cylinders with a height of $h = \SI{400}{nm}$ embedded in a medium with relative permittivity $\varepsilon_\text{h}=2.25$; the radius $r$ and lattice constant $L = 3r$ are varied. The array is illuminated at normal incidence. In (b) the connecting rods have a radius of $\SI{100}{nm}$. (c) Enhancement of optical chirality density in the surrounding medium for cylinders with a radius of $r=\SI{380}{nm}$ at a frequency of $f = \SI{136.22}{THz}$. (d) Volume averaged enhancement of optical chirality density $\langle\hat{C}\rangle$ in the surrounding medium for a left circularly polarized illumination.}
\label{Fig3}
\end{figure}
After the concept has been proven, we now advance towards a more practical structure for a CD enhancement experiment. 
\paragraph{Helicity Preservation and Optical Chirality Enhancement of Periodic Arrays of Cylinders:}
Making accessible meaningful volumes of chiral analytes requires a larger interaction volume between the structure and the molecules. Therefore, identical cylinders are arranged periodically in a square lattice with lattice constant $L=3r$. The array is illuminated at normal incidence with circularly polarized plane-waves. The relative permittivity of the surrounding medium is again $\varepsilon_\text{h}=$2.25. The use of a substrate is avoided, allowing both sides of the structure to be accessed by the molecules. The response of the array is a combination of the response of the cylinders and the electromagnetic coupling between them. Further parameter tuning is necessary. It can be done by maximizing the power transmissivity $T$ of the array, and the degree of helicity preservation is a simple function of the power transmissivity. The $C_4$ rotational symmetry of the structure forces helicity preservation (flip) on the transmitted (reflected) plane-wave \cite{FerCor2013c}. Additionally, in this operational regime, there is only the main diffraction order, and the structure and the surrounding medium are lossless. It follows that $1-T$ is equal to the reflected power and is also equal to the total power of changed helicity. In the right panel of Fig. \ref{Fig3}(a), the maximum is $T=1$ within numerical precision, which amount to zero conversion of helicity to numerical precision. The maximum is achieved for an array of cylinders with radius $r=427.6$ nm, at $f=\SI{124.5}{THz}$. We attribute this improvement over the isolated cylinder to the extra degree of freedom provided by the array pitch, and the reduced number of scattering directions allowed by the array as compared to the isolated cylinder. In order to make the structure manufacturable, the cylinders are connected with smaller rods [see left hand side of Fig.\,\ref{Fig3} (b)]. A comparison of the right hand sides of Figs.\,\ref{Fig3} (a) and (b) shows that the effect of the connecting rods is to slightly change the transmission behavior and the parameter regime in where the desired behavior occurs. It is now observed to be $r=\SI{380}{nm}$ and $f=\SI{136.22}{THz}$, where the transmission is $0.9931$. It follows that $1-0.9931=0.0069$ is the total power of changed helicity, confirming that, while there is a price to pay for the presence of the connecting rods, helicity is preserved to a large degree at the operational point. The phase of the transmitted field (not shown) is shifted by almost exactly $\pi$ with respect to the incident field at this frequency. This amounts to a scattering amplitude of almost exactly $-2$, which is the highest possible for a passive array, and indicates a strong light-matter interaction. We compute the optical chirality enhancement as $\hat{C}=C^+/|C_\text{CPL}|$, where $|C_\text{CPL}|=\epsilon_{\text{h}}\omega/(2c)|E_0|^2$ is the absolute value of the optical chirality density for the circularly polarized plane-wave of amplitude $E_0$ used as illumination. The optical chirality enhancement $\hat{C}=C^+/|C_\text{CPL}|$ is integrated across the surrounding medium in the computational domain and divided by its volume to get a measure for the volume averaged chirality enhancement:
\begin{align}
\langle\hat{C}\rangle=\frac{1}{V}\iiint \frac{C^+(\rr)}{|C_\text{CPL}|}\text{d}V.
\end{align}
The volume of the surrounding medium is given by $V=3r \times 3r \times (h+\SI{400}{nm})-V_\text{s}$, where $V_\text{s}$ is the volume of the silicon structure. The result of this computation is shown in Fig.\,\ref{Fig3} (d). A maximum in the volume averaged chirality enhancement above 6.5 is observed at the optimal frequency. The corresponding spatial distribution of $\hat{C}$ can be seen in Fig.\,\ref{Fig3} (c). The enhancement of optical chirality in the system reflects the \textit{simultaneous} enhancement of the near-field intensity due a strong interaction and the low conversion of the helicity of the illumination. We note that the enhancement of optical chirality depends on the total volume, and hence on the choice of the overall height, currently $h+\SI{400}{nm}$. We will analyze this dependence below. 

\begin{figure}[h!]
\includegraphics[width=\textwidth]{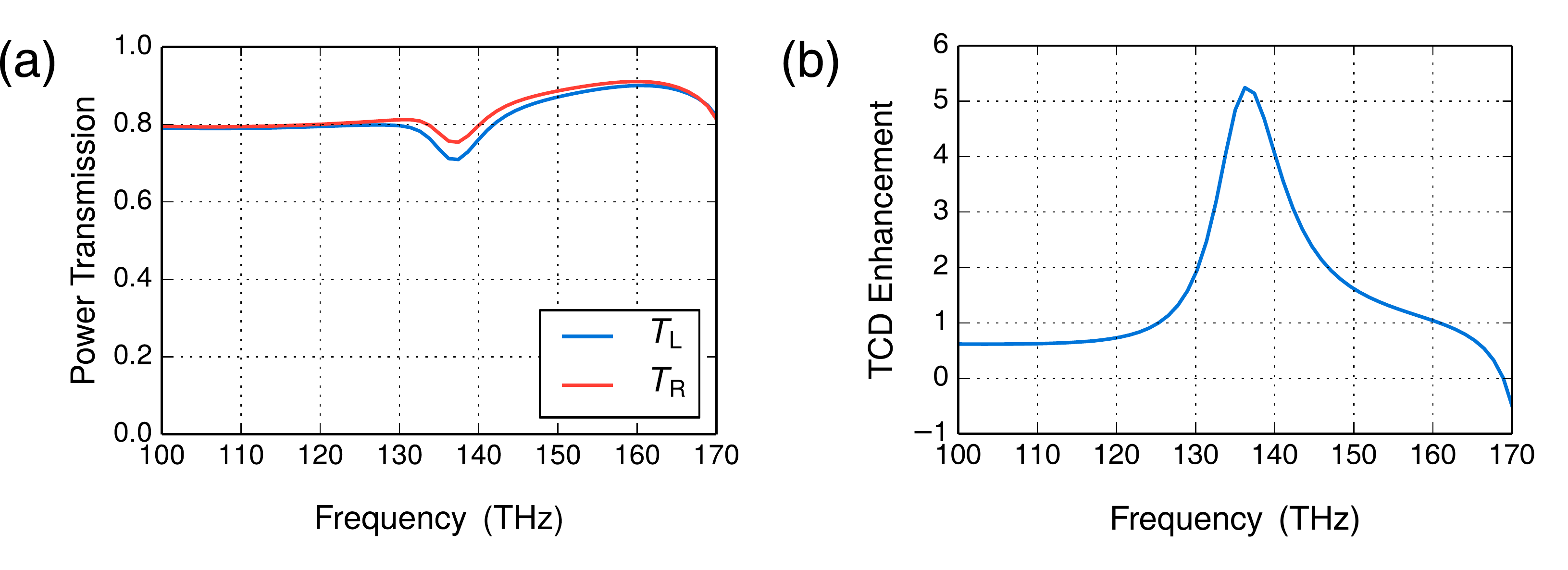}
	\caption{(a) Power transmission of a planar cylinder array embedded in an absorbing Pasteur medium illuminated with left and right circularly polarized plane-waves. $T_{\text{L}}$ and $T_{\text{R}}$ show the transmitted power. (b) Corresponding enhancement in Transmission Circular Dichroism (TCD) $\Delta T/\Delta T_\text{ref}$ due to the array.}
\label{Fig4}
\end{figure}

\begin{figure}[h!]
\includegraphics[height=8cm]{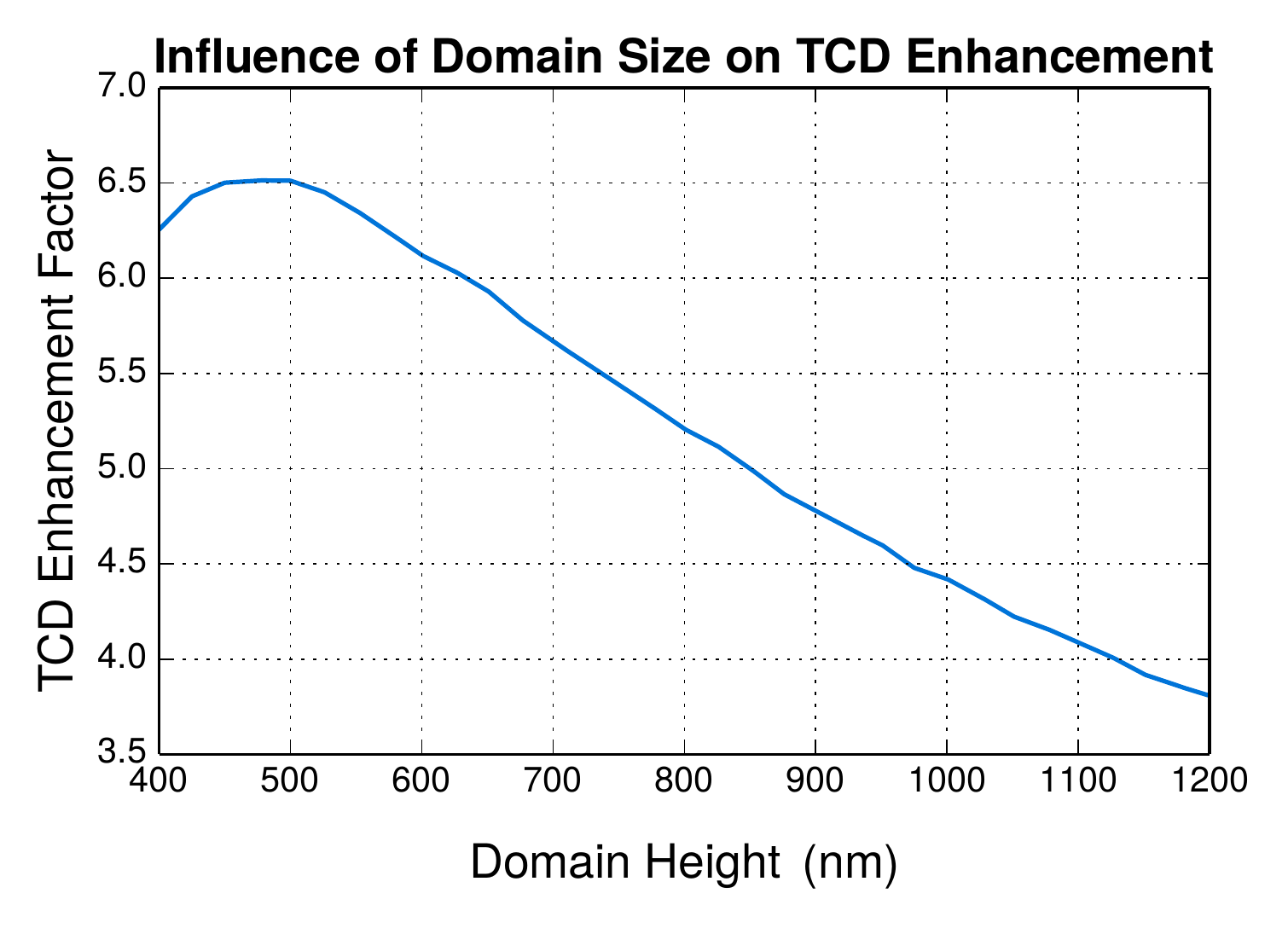}
	\caption{Transmission CD enhancement as a function of the height of the integration volume at the optimal frequency $f=\SI{136.22}{THz}$.}
\label{Fig:TCDDomain}
\end{figure}

\paragraph{Enhancement of the CD Signal:} Finally, the surrounding medium of the array is replaced by a non-dispersive, absorbing Pasteur medium, described by the constitutive relations
\begin{align}\nonumber
	\label{eq:cons}
\textbf{D}&=\epsilon_0\varepsilon_\text{r}\textbf{E}+\text{i}\frac{\kappa}{c_0}\textbf{H},\\
\textbf{B}&=\mu_0\mu_\text{r}\textbf{H}-\text{i}\frac{\kappa}{c_0}\textbf{E}.
\end{align}
The considered medium is characterized by $\varepsilon_\text{r}=2.25+\text{i}0.03$, $\mu_\text{r}=1+\text{i}4\times10^{-5}$, and $\kappa=\text{i}0.001$. The imaginary parts of the permittivity and permeability simulate the achiral absorption of the molecule, and the imaginary part of $\kappa$ the chiral absorption. We note that both permittivity and permeability must have an imaginary part when $\kappa$ has one, since to ensure a lossy medium the following inequality must be met\cite[Eq. 3.219]{Lindell1992}: $(\kappa{''})^2<\varepsilon_\text{r}{''}\mu_\text{r}{''}$. The medium parameters in the constitutive relations of Eq. (\ref{eq:cons}) are assumed to be frequency independent, which would not be the case in a solution of chiral molecules. This simplifying choice allows us to gauge the ``bare'' CD enhancing ability of the system, without the effects of the possible interplay between structural resonances and molecular resonances, which we will consider in future work. The array is illuminated by plane-waves of left- and right- handed circular polarization and, in each case, the transmitted power is computed by integrating the
$z$-component of the Poynting vector on a plane parallel to the array and located in the side opposite to the illumination source. The results of the calculation are given in Fig.\,\ref{Fig4} (a). The power transmissions $T_{\text{L}}$ and $T_{\text{R}}$ show dips at the helicity preserving resonances of the cylinder array. The $T_{\text{L}}$ dip is more pronounced than the $T_{\text{R}}$ one, which is consistent with the fact that the considered Pasteur medium preferentially absorbs the positive helicity. We now compare the calculated differential signal $\Delta T=T_\text{L}-T_\text{R}$ to a reference differential signal of the bare Pasteur medium $\Delta T_\text{ref}=T_\text{L,ref}-T_\text{R,ref}$. For this measurement, the setup is kept as it is but the embedded silicon structure is replaced by more Pasteur medium. We define the transmission circular dichroism (TCD) enhancement of the array as:
\begin{align}
\hat{\text{CD}}_\text{T}=\frac{\Delta T}{\Delta T_\text{ref}}.
\end{align}
The TCD enhancement for the considered cylinder array is shown in Fig.\,\ref{Fig4} (b). A TCD enhancement slightly above 5 is observed at the frequency featuring maximum helicity preservation. Reassuringly, the shape of the TCD enhancement curve shows excellent agreement with the volume averaged chirality enhancement in Fig.\,\ref{Fig3} (d), as predicted by Eq.\,(\ref{eq:dadabartext}). It is important to appreciate that the enhancement depends on the considered volume. In our case, the total volume is controlled by the height of the integration domain, which is 800 nm [see Fig. \ref{Fig3}(c)]. As such height increases, the volumetric size of the regions near the array that experience the most enhanced chirality density becomes relatively smaller with respect to the total volume. In the limit, the enhancement will converge to one. Conversely, by making the height smaller, the enhancement may grow. Figure \ref{Fig:TCDDomain} shows the TCD enhancement as a function of the height of the integration volume at the optimal frequency. It is interesting to see that when the height is just enough to fit the cylinders, the sides of the connected cylinders still provide a significant enhancement, even though smaller than the maximum one. 

It is also important to point out that the enhancement should be independent of the actual value of $\kappa''$. This can be seen by considering Eq. (\ref{eq:dadabartext}) or Eq. (\ref{eq:CDs}) for the cases with and without the structure. The chiral molecular absorption parameter $\alpha_{\text{em}}'$ will always cancel out in a ratio between the CD signals of those two cases. We have checked this in additional simulations, which we include in the Supporting Information. It should be noted, though, that for decreasing values of $|\kappa''|$, an ever finer mesh has to be used for reducing the mirror symmetry breaking that is artificially introduced by meshing the cylinders, which effectively breaks the achirality requirement. Alternatively, a mirror symmetric mesh can very much alleviate the problem \cite{Lee2018}.

Before concluding, we briefly discuss some forward looking points. In our design strategy, the gains come from the strong \textit{and} helicity preserving response of the structures to the incident illuminations. In the case of cylinders, and considering that core-shell spherical designs improve upon the helicity preservation properties of homogeneous spheres \cite{Rahimzadegan2017}, coated cylinders seem a plausible way to improve upon the helicity preservation properties of homogeneous cylinders. Additionally, coated cylinders may also allow to enhance the response strength by aligning together other multipolar resonances beyond the dipoles \cite{Ruan2010}. Another possible way to improve the performance is the optimization of the incident beams constrained to keeping their required relationship through the space-inversion symmetries of the structure. Finally, a set of different arrays could be used to obtain enhancements centered at different frequencies, and may allow to enhance the CD signal across wider bands.

\section*{Conclusion}
We have worked out three requirements for the design of structures that lead to the enhancement of the CD signal of molecular solutions. First: {\em Achirality of the structure plus a corresponding relationship between the two illuminating beams} ensures that the CD signal only depends on the chiral absorption properties of the molecules. Second: If the first requirement is met, {\em helicity preservation} maximizes the gain for a given field enhancing ability of the structure. Third: When the two previous requirements are met, a {\em strong electromagnetic response} will directly enhance the CD signal. We have also designed an exemplary structure following these requirements. We have shown that the aspect ratio of silicon cylinders can be tuned to meet the three requirements at a particular frequency under on-axis plane-wave illumination. Full-wave numerical calculations have shown a several-fold volume-average enhancement of the transmission CD signal for a planar array of connected cylinders. The enhancement is between 6.5 and 3.75 for interaction lengths between 1.25 and 3 times the thickness of the array. Compared to the case where the array is absent, the enhancement would provide the same several-fold reduction of the measurement integration time for a given target signal to noise ratio, or that same several-fold increase in the signal to noise ratio for a fixed integration time. Finally, we have highlighted the usefulness of symmetries and conservation laws, as well as the Riemann-Silberstein representation of the electromagnetic field in the derivation of the requirements and in the design process of the exemplary structure.

\section*{Methods}
Except for the calculations involving the Pasteur medium, all the computations in the paper have been performed with \textit{JCMSuite}. In particular, \textit{JCMSuite} has the built-in ability to compute the multipolar decomposition of the scattered fields in the electric/magnetic basis \cite{Garcia2018}. The calculations involving the Pasteur medium have been performed in \textit{COMSOL} through a modification of its Wave Optics Module to implement the chiral constitutive relations. The details of the implementation are provided in the Supporting Information.

\suppinfo
S1: Derivation of design requirements for CD enhancing structures. S2: Implementation of chiral constitutive relations. S3: Influence of the discretization mesh. 
\begin{acknowledgement}
	We acknowledge support by the Helmholtz Association and KIT via the
	program Science and Technology of Nanosystems (STN) and the Virtual
	Materials Design (VIRTMAT) project. We also thank the Excellence
	Cluster 3D Matter Made to Order for support. We are grateful to the
	company \textit{JCMwave} for their free provision of the FEM Maxwell
	solver \textit{JCMsuite}, with which many of the simulations in this
	work have been performed.  We also warmly thank Mr. Aso
	Rahimzadegan, Mr. Andreas Vetter, and Mr. Philipp Gutsche for useful
	discussions.
\end{acknowledgement}
\appendix
\renewcommand{\thesection}{S\arabic{section}}
\section{Achirality and helicity preservation requirements}
In this appendix, we consider a material structure which is to be used for enhancing the CD signal, and show that:
\begin{enumerate}
	\item Achirality of the structure, and an appropriate relation between the two sequentially incident beams of pure helicity {\em guarantees} that the CD signal will only depend on the chiral absorption properties of the molecules, and not on their achiral (electric and magnetic) absorption properties.
	\item Under the conditions in 1), and given a fixed ability of the structure to enhance the field , helicity preservation maximizes the CD enhancement.
\end{enumerate}

An object is chiral when it lacks invariance under parity, all mirror reflections, and all rotation reflection operations \cite[Chap. 2.6]{Bishop1993}$^,$\cite[p. 26]{Barron2004}. All these transformations can be written generically as the composition of a rotation $R$ and the parity operation $\Pi$ \cite[chap. 2.3]{Bishop1993}. The rotation is of $2\pi/m$ radians about some axis, where $m$ is a positive integer. For $m=1$, we obviously recover the parity operation. For $m=2$ we obtain a mirror reflection about the plane perpendicular to the axis of rotation. For $m>2$ we obtain the rotation-reflection operations. We denote by $M$ such generic space-inversion operation, and start the proof of the two statements above by elucidating the effect of $M$ on the Rienmann-Silberstein representation of electromagnetic fields $\mathbf{G}_\pm(\rr)=\EE\pm iZ\HH$. To such end, we apply $M$ to the electric and magnetic fields:
\begin{equation}
\label{eq:meh}
	\begin{split}
		M\lb\EE\rb=&R\lb\Pi\lb\EE\rb\rb=R\lb-\EEmr\rb=-\lb R_3\mathbf{E}\rb\lb-R_3^{-1}\rr\rb\\
		M\lb\HH\rb=&R\lb\Pi\lb\HH\rb\rb=R\lb\HHmr\rb=\lb R_3\mathbf{H}\rb\lb-R_3^{-1}\rr\rb,
\end{split}
\end{equation}
where the second equality in each line follows from the transformations properties of electric and magnetic fields under parity \cite[Chap. 6.10.B]{Jackson1998}, and the third one from their transformations under rotations\cite[Eq. 2.14]{Zambrana2015}. $R_3$ is the 3$\times$3 matrix that represents the rotation operation $R$ when it acts on 3-vectors. Putting words to Eq. (\ref{eq:meh}): When the transformation $M$ acts on the magnetic field at point $\rr$, it sends it to the image point $\bar{\rr}=-R_3^{-1}\rr$, and, additionally, rotates the components of the 3-vector $\mathbf{H}$. Similarly for the electric field, except that there is an additional overall minus sign. This minus sign causes the expected helicity change inherent to any space-inversion. Indeed, using Eq. (\ref{eq:meh}) it follows that:
\begin{equation}
\label{eq:mrs}
	\begin{split}
		&M\lb\mathbf{G}_\pm(\rr)\rb=M\lb\EE\rb\pm \text{i}Z M\lb\HH\rb\equaldueto{\text{Eq. (\ref{eq:meh})}}-\lb R_3\mathbf{E}\rb\lb-R_3^{-1}\rr\rb\pm \text{i}Z\lb R_3\mathbf{H}\rb\lb-R_3^{-1}\rr\rb=\\
		&-\left[\lb R_3\mathbf{E}\rb\lb-R_3^{-1}\rr\rb\mp \text{i}Z\lb R_3\mathbf{H}\rb\lb-R_3^{-1}\rr\rb\right]=-\lb R_3\mathbf{G}_{\mp}\rb\lb-R_3^{-1}\rr\rb.
		\end{split}
\end{equation}
Namely, the positive(negative) helicity component of the transformed field at the image point $\bar{\rr}=-R_3^{-1}\rr$ is equal to a rotated version of the negative(positive) helicity component at point $\rr$.

We will now consider a CD setup where the molecules are around a material structure which is illuminated sequentially by two different electromagnetic beams of pure and opposite helicity. The effect of the structure on a general electromagnetic field containing both helicity components can be modeled by a linear operator $S$ which takes the incident field and produces the total field. In an abstract notation, we write
\begin{equation}
	\lb\Gop,\Gom\rb=S\lb\Gip,\Gim\rb.
\end{equation}
We consider a first incident field of pure helicity, for example positive, and write
\begin{equation}
	\label{eq:sp}
\lb\Gopp,\Gomp\rb=S\lb\Gip,0\rb,
\end{equation}
where the $+$ superscript in $\lb\Gopp,\Gomp\rb$ denotes that the fields result from illuminating the structure with the field of positive helicity. We now assume that $S$ has at least one space-inversion symmetry $M$ which leaves it invariant, and use $M$ to transform all the objects in Eq. (\ref{eq:sp}). As per our assumption, $S$ does not change, and we obtain:
\begin{equation}
	\label{eq:lr}
	M\lb\Gopp,\Gomp\rb=S\lb M\lb\Gip,0\rb\rb\equaldueto{\text{Eq. (\ref{eq:mrs})}}S\lb0,{\Gim}\rb=\lb\Gopm,\Gomm\rb.
\end{equation}
where the statement $M\lb\Gip,0\rb=\lb0,{\Gim}\rb$ means that, when the initial field of positive helicity $\Gip$ is converted by the $M$ operator, one obtains a new field, which is of opposite helicity, and we denote it by $\Gim$. The point-wise relationship between the two fields is given by Eq. (\ref{eq:mrs}). The leftmost and rightmost parts of Eq. (\ref{eq:lr}) say that the response of the structure to the new incident field of negative helicity ${\Gim}$ can be obtained by using $M$ to transform the response of the structure to the first incident field of positive helicity.  Using this, Eqs. (\ref{eq:mrs}) and (\ref{eq:lr}), and denoting by $\mathbf{G}_\lambda^{\mathrm{tot},\bar{\lambda}}(\rr)$ the helicity $\lambda$ component of the total field at point $\rr$ due to the incident beam of helicity $\bar{\lambda}$, we can write:
\begin{equation}
	\label{eq:key}
		\Gmmt=-\Gppbr,\text{ and } \Gpmt=-\Gmpbr.\\ 
\end{equation}
At this point, we exclude the rotation-reflection operations, which we will include again at the end. The remaining operations, i.e., parity and mirror reflections, meet $M^2=I$, where $I$ is the identity. If we apply $M$ again to Eq. (\ref{eq:lr}), and since $M^2=I$, we will obtain that the response of the structure to the first incident field of positive helicity ${\Gip}$ can be obtained by using $M$ to transform the response of the structure to the second incident field of negative helicity, which will lead to expressions similar to Eq. (\ref{eq:key}):
\begin{equation}
	\label{eq:key2}
		\Gppt=-\Gmmbr,\text{ and } \Gmpt=-\Gpmbr. 
\end{equation}
Let us now examine the CD measurement where $\lb\Gip,0\rb$ and $\lb0,{\Gim}\rb$ are the two sequentially incident beams. We consider the pointwise differential absorption formula in Eq. (\ref{eq:datext}) of the main text. From this point on, all fields are total fields and we drop the $^{\mathrm{tot}}$ superscript. 
\begin{equation}
	\label{eq:da}
	\begin{split}
		\Delta A (\rr)&=A^+(\rr)-A^-(\rr)=\\
		&\frac{\omega}{8}\lb |\Gpp|^2+|\Gmp|^2-|\Gmm|^2-|\Gpm|^2\rb\lb\alpha_{\text{ee}}''+\epsilon\alpha_{\text{mm}}''\rb\\
		&+\frac{\omega}{4}\lb \mathrm{Re}\{{\Gpp}^\dagger\Gmp-{\Gpm}^\dagger\Gmm\}\rb\lb\alpha_{\text{ee}}''-\epsilon\alpha_{\text{mm}}''\rb\\
		&-\frac{\omega}{4Z}\lb |\Gpp|^2-|\Gmp|^2-|\Gpm|^2+|\Gmm|^2\rb\alpha_{\text{em}}'. 
	\end{split}
\end{equation}
The CD signal is the integral of the differential absorption in Eq. (\ref{eq:da}) over the volume occupied by the molecules. This means that the differential absorptions at image pairs $\rr$ and $\bar{\rr}$ will be added together. If we consider such sum $\Delta A (\rr)+\Delta A (\bar{\rr})$, and use the equalities in Eqs. (\ref{eq:key})-(\ref{eq:key2}), it is a matter of straightforward algebraic manipulations (written later) to show that all the terms that multiply the achiral absorption properties of the molecule, which depend on $\alpha_{\text{ee}}''$ and $\alpha_{\text{mm}}''$, cancel out. The resulting CD signal can be written: 
\begin{equation}
	\label{eq:dadabar}
	CD=\int_V d\rr\left[ |\Gpp|^2+|\Gmm|^2-\lb|\Gmp|^2+|\Gpm|^2\rb\right]\left[-\frac{\alpha_{\text{em}}'\omega}{4Z}\right].
\end{equation}
Therefore, the cancellation of the terms that depend on the achiral absorption properties of the molecules $(\alpha_{\text{ee}}'',\alpha_{\text{mm}}'')$ is {\em guaranteed} by using two incident fields that transform into each other with a parity or mirror reflection operation that is a symmetry of the structure $S$. $S$ is hence achiral.

Additionally, Eq. (\ref{eq:dadabar}) can be used to show that {\em for a given total field enhancement, a helicity preserving structure, which will force $\Gmp=\Gpm=0$ for all $\rr$, will maximize the CD enhancement.} This can be seen as follows. The total field intensity is proportional to
\begin{equation}
	\int_V d\rr \left[|\Gpp|^2+|\Gmm|^2+\lb|\Gmp|^2+|\Gpm|^2\rb\right].
\end{equation}
When this quantity is fixed, one way to maximize Eq. (\ref{eq:dadabar})
is to have $|\Gmp|^2=|\Gpm|^2=0$, that is, to have helicity
preservation. The other way, which will produce a CD signal of the
opposite sign, would be to have $|\Gpp|=|\Gmm|=0$. This means that the
structure generates a scattered field which is minus the incident field,
plus a field of changed helicity. We deem this second route to be more challenging and do not consider it further. 

We now write the algebra that reduces Eq. (\ref{eq:da}) to the integrand in Eq. (\ref{eq:dadabar}) when the conditions in Eqs. (\ref{eq:key})-(\ref{eq:key2}) are met. It follows from Eq. (\ref{eq:key}) that: 
{
\begin{equation}
	\label{eq:small}
	\begin{split}
		|\Gmmtnt|^2&=\lb-\Gppbrnt\rb^\dagger\lb-\Gppbrnt\rb=\lb\mathbf{G}_+^{+}(\bar{\rr})\rb^\dagger R_3^\dagger R_3\mathbf{G}_+^{+}(\bar{\rr})=|\mathbf{G}_+^{+}(\bar{\rr})|^2,\\
		|\Gpmtnt|^2&=\lb-\Gmpbrnt\rb^\dagger\lb-\Gmpbrnt\rb=\lb\mathbf{G}_-^{+}(\bar{\rr})\rb^\dagger R_3^\dagger R_3\mathbf{G}_-^{+}(\bar{\rr})=|\mathbf{G}_-^{+}(\bar{\rr})|^2,
				\end{split}
\end{equation}
}
where the third equalities in each line follow from the unitary character of any rotation $R_3$. Similarly, it follows from Eq. (\ref{eq:key2}) that $|\Gpp|^2=|\Gmmb|^2$, and $|\Gmp|^2=|\Gpmb|^2$. Using all these relationships, it is straightforward to see that, when we add $\Delta A (\rr)+\Delta A (\bar{\rr})$, the terms corresponding to the second line in Eq. (\ref{eq:da}) will cancel, and that those corresponding to the fourth line will result in Eq. (\ref{eq:dadabar}). The cancellation of the terms corresponding to the third line of Eq. (\ref{eq:da}) is a bit more involved. First, we use again Eqs. (\ref{eq:key}) and (\ref{eq:key2}) and steps like those in Eq. (\ref{eq:small}) to show that
\begin{equation}
		\Gpm^\dagger\Gmm=\mathbf{G}_-^{+}(\bar{\rr})^\dagger\mathbf{G}_+^{+}(\bar{\rr}),\text{ and } \Gmp^\dagger\Gpp=\mathbf{G}_+^{-}(\bar{\rr})^\dagger\mathbf{G}_-^{-}(\bar{\rr}),
\end{equation}
which we use in the relevant sum
\begin{equation}
\mathrm{Re}\{{\Gpp}^\dagger\Gmp-{\Gpm}^\dagger\Gmm\}+\mathrm{Re}\{{\Gppb}^\dagger\Gmpb-{\Gpmb}^\dagger\Gmmb\}
\end{equation}
to get
\begin{equation}
		\mathrm{Re}\{{\Gpp}^\dagger\Gmp-\mathbf{G}_-^{+}(\rr)^\dagger\mathbf{G}_+^{+}(\rr)\}+\mathrm{Re}\{{\Gppb}^\dagger\Gmpb-{\Gmpb}^\dagger\Gppb\},\\
\end{equation}
where both terms inside the real parts are of the form $\mathrm{Re}\{a-a^*\}$, which is equal to zero.

We can also use the equalities 
\begin{equation}
	\label{eq:equalities}
	|\Gmmtnt|^2=|\mathbf{G}_+^{+}(\bar{\rr})|^2,\ |\Gpmtnt|^2=|\mathbf{G}_-^{+}(\bar{\rr})|^2,\ |\Gpp|^2=|\Gmmb|^2,\text{ and }|\Gmp|^2=|\Gpmb|^2,
\end{equation}
to work on the expression of the CD signal in Eq. (\ref{eq:dadabartext}) of the main text, which already assumes that the structure has an inversion symmetry and the illuminations are related accordingly:
\begin{equation}
	\label{eq:cisgood}
	\begin{split}
		\text{CD}&=-2\alpha_{\text{em}}'c^2\int_V d\rr \left[ C^+(\rr)-C^-(\rr)\right]=\\
		&-\frac{\alpha_{\text{em}}'\omega}{4Z}\int_V d\rr \left[ |\Gpp|^2-|\Gmp|^2-\lb |\Gpm|^2-|\Gmm|^2\rb\right]\equaldueto{\text{Eq. (\ref{eq:equalities})}}\\
		&-\frac{\alpha_{\text{em}}'\omega}{2Z}\int_V d\rr \left[ |\Gpp|^2-|\Gmp|^2\right]\\
		&\equaldueto{\text{Eq. (\ref{eq:equalities})}}-4\alpha_{\text{em}}'c^2\int_V d\rr \ C^+(\rr)\equaldueto{\text{Eq. (\ref{eq:equalities})}}4\alpha_{\text{em}}'c^2\int_V d\rr \ C^-(\rr).
	\end{split}
\end{equation}
The last line which shows that $C^+(\rr)$[$C^-(\rr)$] and its integrated value are a good indicator for the CD signal under the mentioned conditions.

Finally, we consider the excluded rotation-reflection symmetries. The derivation to show that the same results apply to them is identical to the one written down here, except that, instead of a single image point $\bar{\rr}$, there would be a larger number of them $\bar{\rr}_l$. For a rotation-reflection symmetry containing a rotation by $2\pi/m$ for $m>2$, the total number of points including the original point $\rr$ would be $m$ if $m$ is even and $2m$ if m is odd. This condition follows from the smallest value of $k$ that meets the equation $M^k=I$

\section{Implementation of chiral constitutive relations}
\textit{COMSOL} allows the user to modify the underlying equations that are used for the calculations. This provides the freedom to implement bi-isotropic constitutive relations:
\begin{align}\nonumber
{{\mathbf{D}}}=\epsilon_0\epsilon_\text{r}{{\mathbf{E}}}+\frac{\chi-\mathrm{i}\kappa}{c_0}{{\mathbf{H}}}\,, \qquad {{\mathbf{H}}}=\frac{1}{\mu_0\mu_\text{r}}\Big({{\mathbf{B}}}-\frac{\chi+\mathrm{i}\kappa}{c_0}{{\mathbf{E}}}\Big)\,.
\end{align}
Again, $\kappa$ is the chirality parameter of the medium. For the sake of generality we also include the parameter $\chi$ that is used to describe a cophasal magnetoelectric effect. Media that exhibit $\chi\neq 0$ are nonreciprocal.

The modifications are applied to the Wave Optics Module used in frequency domain. One enables the "Equation View" gaining access to the underlying equations and selecting "Electromagnetic Waves, Frequency Domain (ewfd)" and finally entering the equation view of "Wave Equation, Electric 1". The displacement vector ${\mathbf{D}}$ is related to the polarization vector ${\mathbf{P}}$ via, ${{\mathbf{D}}}=\epsilon_0{{\mathbf{E}}}+{{\mathbf{P}}}$ with ${{\mathbf{P}}}=\epsilon_0{{\epsilon}}{{\mathbf{E}}}$. The equation for the displacement field components is modified as

\noindent\textsf{ewfd.Dx : epsilon0$\_$const*ewfd.Ex+ewfd.Px\textcolor{red}{+(chi-i*kappa)/c0*ewfd.Hx}}\\
\noindent\textsf{ewfd.Dy : epsilon0$\_$const*ewfd.Ey+ewfd.Py\textcolor{red}{+(chi-i*kappa)/c0*ewfd.Hy}}\\
\noindent\textsf{ewfd.Dz : epsilon0$\_$const*ewfd.Ez+ewfd.Pz\textcolor{red}{+(chi-i*kappa)/c0*ewfd.Hz}$\,$.}

Here the red parts show the applied modifications. The constitutive relations for the components of ${\mathbf{H}}$ and $\mathbf{d}{\mathbf{H}}/\mathbf{d}t$ are modified as:\\
	\noindent \textsf{ewfd.Hx : (ewfd.murinvxx*\textcolor{red}{(}ewfd.Bx\textcolor{red}{-(chi+i*kappa)/c0*ewfd.Ex)}}+\\\textsf{ewfd.murinvxy*\textcolor{red}{(}ewfd.By\textcolor{red}{-(chi+i*kappa)/c0*ewfd.Ey)}}+\\\textsf{ewfd.murinvxz*\textcolor{red}{(}ewfd.Bz\textcolor{red}{-(chi+i*kappa)/c0*ewfd.Ez)})/mu0\_const}\\
	\textsf{ewfd.Hy : (ewfd.murinvyx*\textcolor{red}{(}ewfd.Bx\textcolor{red}{-(chi+i*kappa)/c0*ewfd.Ex)}}+\\\textsf{ewfd.murinvyy*\textcolor{red}{(}ewfd.By\textcolor{red}{-(chi+i*kappa)/c0*ewfd.Ey)}}+\\\textsf{ewfd.murinvyz*\textcolor{red}{(}ewfd.Bz\textcolor{red}{-(chi+i*kappa)/c0*ewfd.Ez)})/mu0\_const}\\
	\textsf{ewfd.Hz : (ewfd.murinvzx*\textcolor{red}{(}ewfd.Bx\textcolor{red}{-(chi+i*kappa)/c0*ewfd.Ex)}}+\\\textsf{ewfd.murinvzy*\textcolor{red}{(}ewfd.By\textcolor{red}{-(chi+i*kappa)/c0*ewfd.Ey)}}+\\\textsf{ewfd.murinvzz*\textcolor{red}{(}ewfd.Bz\textcolor{red}{-(chi+i*kappa)/c0*ewfd.Ez)})/mu0\_const}\\
\textsf{ewfd.dHdtx : (ewfd.murinvxx*\textcolor{red}{(}ewfd.dBdtx\textcolor{red}{-(chi+i*kappa)/c0*ewfd.iomega*ewfd.Ex)}+\\ ewfd.murinvxy*\textcolor{red}{(}ewfd.dBdty\textcolor{red}{-(chi+i*kappa)/c0*ewfd.iomega*ewfd.Ey)}+ewfd.murinvxz*\\ \textcolor{red}{(}ewfd.dBdtz\textcolor{red}{-(chi+i*kappa)/c0*ewfd.iomega*ewfd.Ez)})/mu0$\_$const}\\
\textsf{ewfd.dHdty : (ewfd.murinvyx*\textcolor{red}{(}ewfd.dBdtx\textcolor{red}{-(chi+i*kappa)/c0*ewfd.iomega*ewfd.Ex)}+\\ ewfd.murinvyy*\textcolor{red}{(}ewfd.dBdty\textcolor{red}{-(chi+i*kappa)/c0*ewfd.iomega*ewfd.Ey)}+ewfd.murinvyz*\\ \textcolor{red}{(}ewfd.dBdtz\textcolor{red}{-(chi+i*kappa)/c0*ewfd.iomega*ewfd.Ez)})/mu0$\_$const}\\
\textsf{ewfd.dHdtz : (ewfd.murinvzx*\textcolor{red}{(}ewfd.dBdtx\textcolor{red}{-(chi+i*kappa)/c0*ewfd.iomega*ewfd.Ex)}+\\ ewfd.murinvzy*\textcolor{red}{(}ewfd.dBdty\textcolor{red}{-(chi+i*kappa)/c0*ewfd.iomega*ewfd.Ey)}+ewfd.murinvzz*\\ \textcolor{red}{(}ewfd.dBdtz\textcolor{red}{-(chi+i*kappa)/c0*ewfd.iomega*ewfd.Ez)})/mu0$\_$const$\,$.}

Before the implementation can be used, the following variables need to be defined:  "\textsf{chi}", "\textsf{kappa}", and "\textsf{c0=1/sqrt(epsilon0$\_$const*mu0$\_$const)}" in "Component 1" $\rightarrow$ "Definitions" $\rightarrow$ "Variables". 

The implementation was tested by comparing the numerically obtained reflection and transmission coefficients of a chiral slab with a thickness of $L=\SI{500}{nm}$ embedded in air with their analytically obtained values\cite{Lindell1994}. We obtained a perfect match.

\section{Influence of the Discretization Mesh}
\begin{figure}[h!]
\includegraphics[width=5.25in]{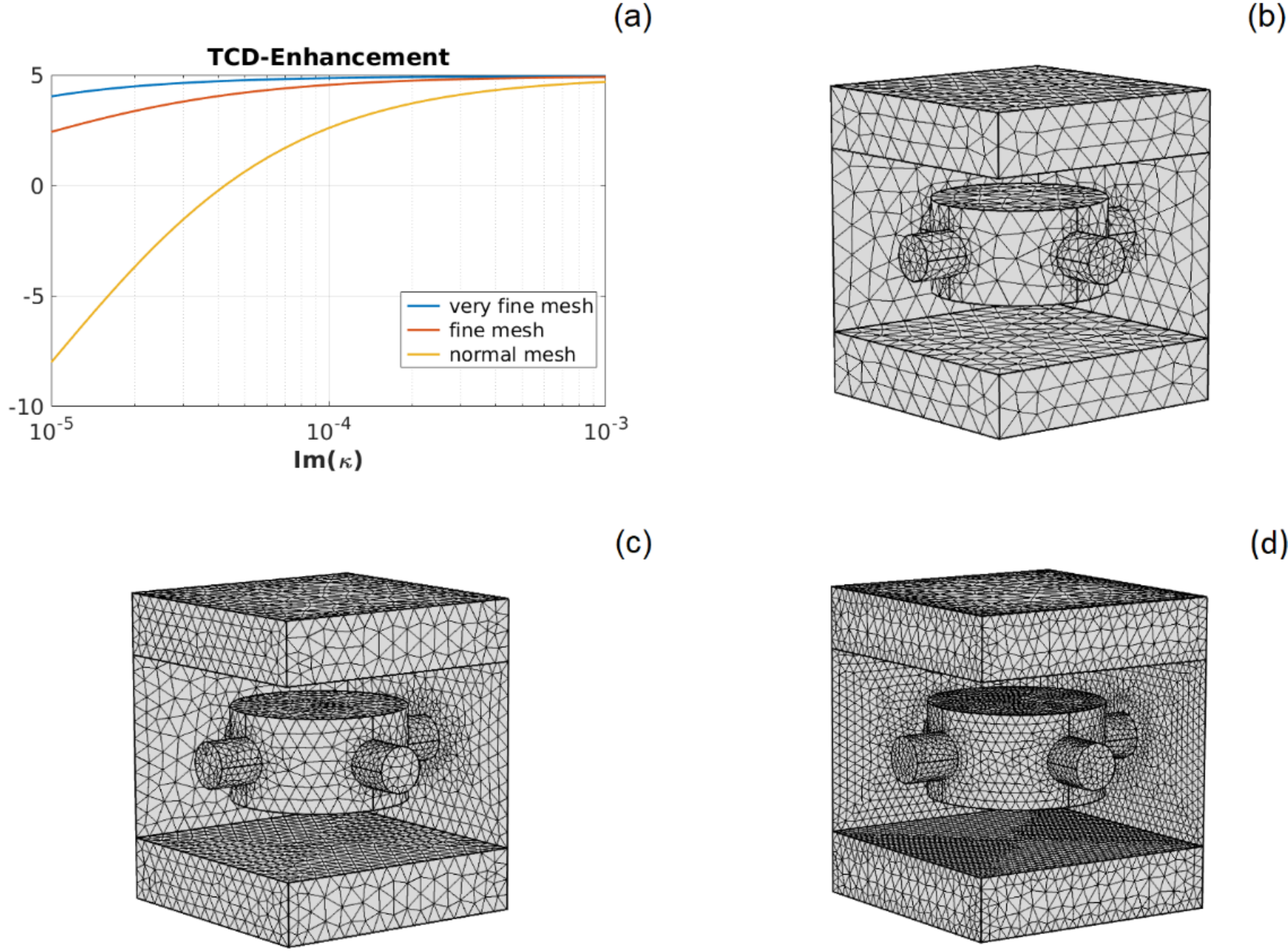}
	\caption{(a) Comparison of the calculated TCD-enhancements $\Delta T/\Delta T_\text{ref}$ for three different meshes at a frequency of $f=\SI{135}{THz}$ and different values of $\kappa''$. (b) Visualization of the "normal mesh", maximum discretization step 133 nm. (c) Visualization of the "fine mesh", maximum discretization step 107 nm. (d) Visualization of the "very fine mesh", maximum discretization step 47 nm. The interpolation used a cubic polynomial in all cases.}
\label{kappacheck}
\end{figure}
The value of the imaginary part of $\kappa$ in the main article is $10^{-3}$, which is larger than for typical chiral solutions. Simulations using values such as $10^{-6}-10^{-4}$ require extremely high numerical accuracy, i.e. they are computationally expensive. As argued in the main text, the CD enhancement does not depend on the actual value of $\kappa''$ for structures and illumination choices that meet the achirality requirement. We performed calculations to verify this statement by fixing the frequency at $f=\SI{135}{THz}$ and varying $\kappa''$ and the mesh densities. The results are shown in Fig. \ref{kappacheck}. For an ordinarily sized mesh as depicted in Fig. \ref{kappacheck} (b) the TCD enhancement decreases for very small $\kappa''$ and even becomes negative. With a finer mesh [cf. Fig.\,\ref{kappacheck} (c)], however, the TCD enhancement stays stable down to values $\kappa''=10^{-4}$ and then starts to decrease. An even finer mesh [cf. Fig.\,\ref{kappacheck} (d)] can keep the TCD enhancement stable for values down to almost $\kappa''=10^{-5}$. With this finest mesh we reached the limit of our available memory resources. We take the continuous improvement with decreasing mesh size and the fact that for the finest mesh the TCD enhancement stays approximately constant over two decades, is consistent with the enhancement factor being independent of the magnitude of $\kappa''$. The fact that, for the mesh depicted in Fig. \ref{kappacheck} (b), the enhancement can change sign indicates a breaking of the theoretical mirror symmetry of the structure due to the mesh. Indeed, the mesh causes the structure to become effectively chiral because, upon space-inversion, the discretization volumes do not overlap with the original ones. Accordingly, this effect diminishes for smaller meshes, which is consistent with the results in \cite{Lee2018}.

 \providecommand{\latin}[1]{#1}
\makeatletter
\providecommand{\doi}
  {\begingroup\let\do\@makeother\dospecials
  \catcode`\{=1 \catcode`\}=2\doi@aux}
\providecommand{\doi@aux}[1]{\endgroup\texttt{#1}}
\makeatother
\providecommand*\mcitethebibliography{\thebibliography}
\csname @ifundefined\endcsname{endmcitethebibliography}
  {\let\endmcitethebibliography\endthebibliography}{}

\end{document}